\documentclass[aps,twocolumn,amsmath,showpacs,floatfix]{revtex4}
\usepackage{mathrsfs,bm,times}
\usepackage{graphicx}
\usepackage{amsmath,amsfonts}
\allowdisplaybreaks

\makeindex

\begin{document}
\title{
Three-dimensional higher-spin Dirac and Weyl dispersions in the
strongly isotropic $\bm{K_4}$ crystal}

\date{November 17, 2016}

\author{Masahisa Tsuchiizu}

\affiliation{Department of Physics, Nagoya University, Nagoya 464-8602, Japan}

\begin{abstract}
We analyze the  electronic structure in the 
three-dimensional (3D) crystal formed by
 the $sp^2$ hybridized orbitals ($K_4$ crystal), 
by the tight-binding approach based on the 
first-principles calculation.
We discover that the bulk Dirac-cone dispersions are realized 
   in the $K_4$ crystal.
In contrast to the graphene,
  the energy dispersions of the Dirac cones 
 are isotropic in 3D and
the pseudospin $S=1$ Dirac cones emerge at the
$\Gamma$ and $H$ points of the bcc Brillouin zone,
where 
 three bands become degenerate and merge at a single point 
  belonging to the $T_2$ irreducible representation.
In addition, the usual $S=1/2$ Dirac cones
emerge at the $P$ point.
By focusing the hoppings between the nearest-neighbor sites, 
 we show an analytic form of the tight-binding
 Hamiltonian with a  $4\times 4 $ matrix, and we
give an explicit derivation of the  $S=1$ and $S=1/2$ 
 Dirac-cone dispersions. 
We also analyze the effect of the spin-orbit coupling 
to examine how the degeneracies at Dirac points are lifted.
At the $S=1$ Dirac points,
the spin-orbit coupling lifts the energy level with sixfold degeneracy
   into two energy levels with two-dimensional $\bar E_2$ and 
 four-dimensional $\bar F$ representations.
Remarkably,
all the dispersions near the $\bar F$ point
show the linear dependence in the momentum with different velocities.
We derive the effective Hamiltonian near the $\bar F$ point and
 find that the band contact point is described by the $S=3/2$
Weyl point.
\end{abstract}

\pacs{71.20.Gj, 31.15.aq, 73.22.Pr}

\maketitle

\section{Introduction}

Electronic structure of graphene has been a subject of intensive
research over the years 
\cite{Ando20021,Abergel:2010jz,Aoki_book,Yamakage_book}, since 
it has been recognized as the most
 exciting material 
 after the discovery of the 
massless Dirac fermions 
\cite{PhysRev.71.622,PhysRev.109.272}.
The massless Dirac fermions 
have been widely recognized in the  condensed-matter systems,
especially in the context of the topological insulators 
\cite{Fu:2007io,Hasan:2010ku}.
In the case of graphene,
the $sp^2$ hybridized orbitals build up the honeycomb crystal, and 
the $\pi$ electrons 
 exhibit the Dirac fermion behavior on it. 
Recently, the 3D analog of the graphene has also attracted 
much attention
\cite{Young:2012kz,Yang:2014ia}.
Even in the 3D diamond structure, the suppression of the density of states 
  has been observed in the valence band
\cite{Chadi:1975aa}
and the possible realization of the three-dimensional (3D) Dirac cone
  has been discussed
\cite{Young:2012kz}.

\begin{figure}[b]
\includegraphics[width=8.5cm,bb=50 99 773 497]{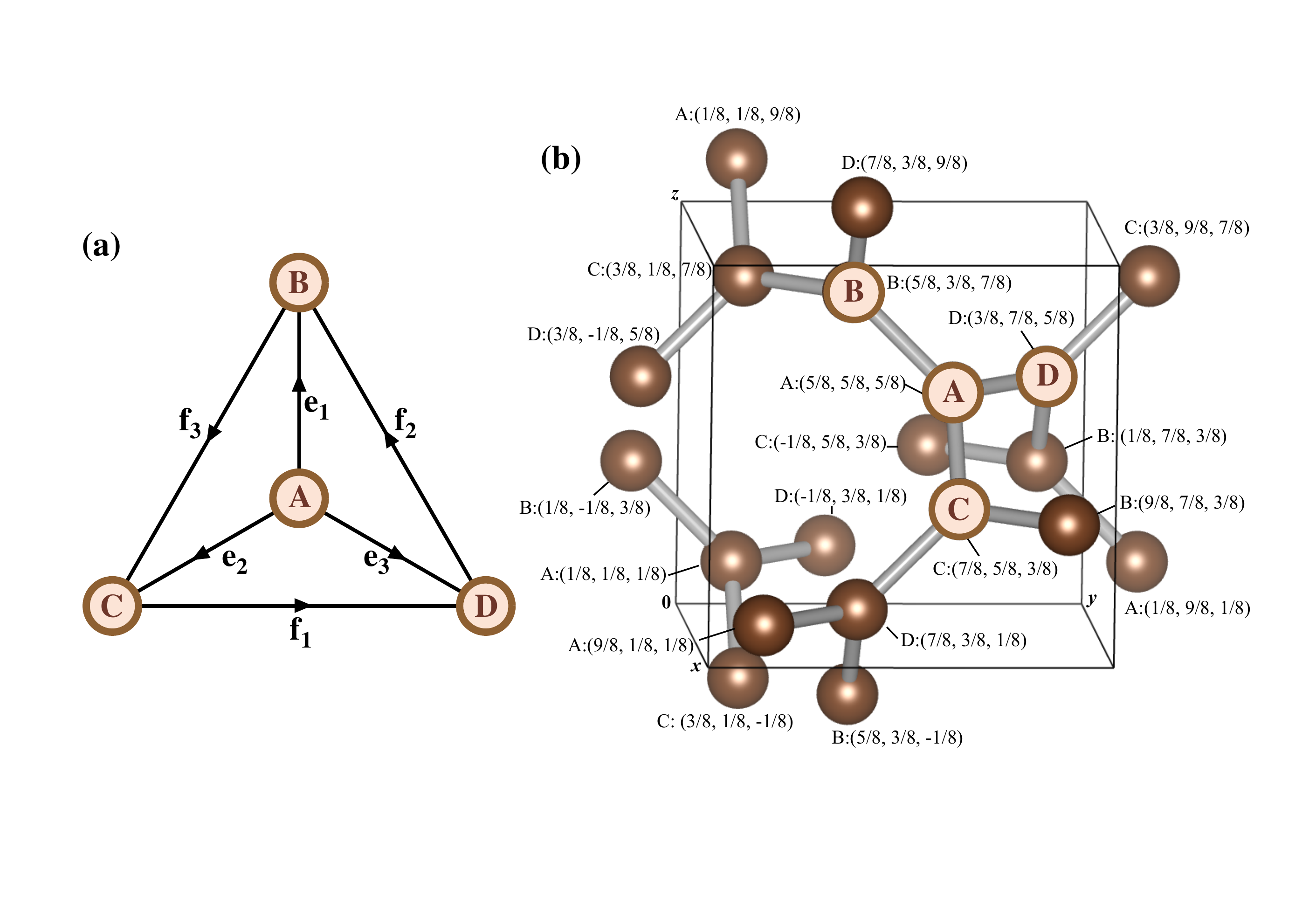}
\caption{
 (a) Complete graph $K_4$ where
$\{A,B,C,D\}$ are the vertices and
$\{e_1,e_2,e_3,f_1,f_2,f_3\}$
are the oriented edges.
(b) The $K_4$ crystal structure.
The numbers in the parentheses represent the fractional 
Cartesian coordinates
  in the cubic conventional unit cell.
There are eight sites in the conventional unit cell.
In the primitive unit cell, there are four sites, 
$A$, $B$, $C$, and $D$.
The coordination number is three and 
the bonds connecting nearest-neighbor sites are represented by the
 vectors given in Eq.\ (\ref{eq:vectors}).
}
\label{fig1}
\end{figure}

From the mathematical point of view, the honeycomb  and 
the diamond crystals have the common properties, called the 
  \textit{strongly isotropic} property
\cite{Sunada:2007tb,Sunada:2012ve}.
The strongly isotropic property
indicates
the property that preserves the crystal net after 
any permutation of bonds with common end point.
The honeycomb structure is the only 2D crystal that possesses the 
strongly isotropic property.
In 3D, there are only two strongly isotropic crystals:
one is the diamond crystal and the other is the $K_4$ crystal.
The K$_4$ crystal is defined 
as the standard realization of the maximal topological crystal over the
   $K_4$ graph \cite{Sunada:2007tb,Sunada:2012ve}. 
Here $K_4$ represents the complete graph with four vertices,
as shown in Fig.\ \ref{fig1}(a), and its crystal structure
is shown in Fig.\ \ref{fig1}(b).
Like the 2D honeycomb crystal,
   the coordination number of the $K_4$ crystal is three.
The synthesis of the $K_4$ crystal in terms of the carbon atoms
(called the $K_4$ carbon) has not been succeeded so far
despite the several theoretical predictions
based on the first-principles calculations
\cite{Rignanese:2008ii,Itoh:2009gs}.
Quite recently, the discovery of the $K_4$ crystal was made
where the constituting unit is a large molecule instead of the carbon atom
\cite{Mizuno:2015ip}.

%====================================================================
%======================================================================

The effect of the spin-orbit coupling (SOC) on the Dirac cones
has been  attracting great interest owing to the discovery of 
the topological insulator   \cite{Kane:2005hl}.
In the case of honeycomb structure,  
 the degeneracy of the Dirac point is lifted 
  and the gap appears.
On the other hand, in the case of diamond crystal, 
  the SOC lifts the degeneracy along the $X$-$W$ line in 
the Brillouin zone \cite{Young:2012kz} except for the $X$ point,
i.e., the two bands touch at the $X$ point.
Due to the presence of the inversion (I) and time-reversal (TR) symmetries, 
each band is doubly degenerate at general $\bm k$ points and thus 
the contact point has fourfold degeneracy.
Thus the band touching point at the $X$ point 
is the Dirac point, which is described by the four-band 
 Dirac Hamiltonian.
If the I or TR symmetry is broken,
  the double degeneracy at general $\bm k$ points is lifted.
In this case, 
  the possibility of the low-symmetry band touchings has been 
    argued
\cite{Halasz:2012ec,Liu:2014gy}, 
 where the contact point has twofold degeneracy.
These touching points are referred to as the Weyl points
\cite{Wan:2011hi,Burkov:2011tu}.
Recently, the experimental realization of the Weyl points 
 has been reported \cite{Lu:2015tp}.

In the present paper,
 we analyze the band structure of the $K_4$
 crystal by the tight-binding approach.
We discover that three-dimensional Dirac-cone dispersions 
are realized in the $K_4$ crystal.
In contrast to the graphene, 
  three bands touch at a single point  on the $\Gamma$ and $H$ points
 of the bcc Brillouin zone, indicating the emergence of pseudospin 
 $S=1$ Dirac cone.
In addition, the usual $S=1/2$ Dirac cones emerge at the 
$P$ point, where the two bands touch at a single point.
Since the I symmetry is broken in the  $K_4$ crystal, 
the energy splitting due to SOC is peculiar.
We show that, 
at the $S=1$ Dirac points,
the SOC lifts the energy level with sixfold degeneracy
   into two energy levels with  two-dimensional $\bar E_2$ and 
four-dimensional $\bar F$ irreducible representations.
Especially, we find that
the dispersion near the $\bar F$ point
is described by the $S=3/2$ Weyl dispersions.

The present paper is organized as follows. 
In Sec.\ \ref{sec:K4crystal}, we recall briefly how the $K_4$ crystal 
can be realized from the 
the $K_4$ graph and how
the  strongly isotropic character is retained.
In Sec.\ \ref{sec:K4carbon},
we construct the tight-binding model for the $K_4$ carbon based on
the first-principles calculation by focusing on the carbon $\pi$ orbital.
In Sec.\ \ref{sec:Dirac},
the tight-binding Hamiltonian 
is analyzed 
and we show explicitly 
how the pseudospin $S=1$ and $S=1/2$  Dirac-cone
 dispersions 
  are derived.
Finally, 
in Sec.\ \ref{sec:SOC},
the effect of the SOC is analyzed.
Section \ref{sec:summary} is devoted to the summary and discussions.

\section{$\bm K_4$ crystal}\label{sec:K4crystal}

The $K_4$ crystal is realized 
as  the maximal topological crystal over the
   $K_4$ graph.
The general arguments 
 to obtain the realization of the crystal from the 
 finite graph are based on the homology group.
Here, we simply follow the algorithm given in Refs.\ 
\cite{Sunada:2007tb} and \cite{Sunada:2012ve} 
 to construct of the $K_4$ crystal, 
without the mathematical details.
In the graph $K_4$ of Fig.\ \ref{fig1}(a), the 
vertices are described by $\{A, B, C, D\}$ and the edges are
$\{e_1,e_2,e_3,f_1,f_2,f_3\}$.
First we consider three closed paths
 $c_1=(e_2,f_1,\bar{e}_3)$,
 $c_2=(e_3,f_2,\bar{e}_1)$, and
 $c_3=(e_1,f_3,\bar{e}_2)$
 in the $K_4$ graph [Fig.\ \ref{fig1}(a)].
The inner product can be introduced by 
$\langle e, e'\rangle = 1$ (if $e'=e$), $-1$ (if $e'=\bar e$), and $0$
(otherwise), where $e,e'=e_i,f_i$.
Then we find 
$|c_1|^2=|c_2|^2=|c_3|^2=3$ and 
$\langle c_i, c_j\rangle = -1 $ ($i\ne j$).
The vectors connecting the nearest-neighbor sites (building blocks)
for the infinite  $K_4$ crystal, $\bm v(e_i)$ and
$\bm v(f_i)$, 
can be constructed 
  by taking $c_1$, $c_2$, $c_3$ as the basis, e.g.,
$\bm v(e_1)=a_1 \bm c_1 + a_2 \bm c_2 + a_3 \bm c_3$ 
($a_{i} \in \mathbb{R}$).
From the relations 
$\langle \bm v(e_1), c_1\rangle = \langle e_1, c_1\rangle = 0 $,
$\langle \bm v(e_1), c_2\rangle = \langle e_1, c_2\rangle = -1$, and
$\langle \bm v(e_1), c_3\rangle = \langle e_1, c_3\rangle = +1$, 
we can obtain $a_1=0$, $a_2=-1/4$, and $a_3=+1/4$, and thus
 $\bm v(e_1)$ is determined as
$\bm v(e_1) = -\bm c_2/4 + \bm c_3/4$.
From this simple calculation,
we can get the following relations:
%===================================================================
\begin{eqnarray}
\bm v(e_1) &=& -\frac{1}{4} \bm c_2 + \frac{1}{4} \bm c_3,
 \nonumber  \\
 \bm v(e_2) &=& -\frac{1}{4} \bm c_3 + \frac{1}{4} \bm c_1,
\nonumber   \\
 \bm v(e_3) &=& -\frac{1}{4} \bm c_1 + \frac{1}{4} \bm c_2,
\label{eq:vectors}
  \\
 \bm v(f_1) &=& +\frac{1}{2} \bm c_1 + \frac{1}{4} \bm c_2 
 + \frac{1}{4} \bm c_3,
\nonumber \\
 \bm v(f_2) &=& +\frac{1}{2} \bm c_2 + \frac{1}{4} \bm c_3 
 +\frac{1}{4}  \bm c_1,
\nonumber   \\
 \bm v(f_3) &=& +\frac{1}{2} \bm c_3+\frac{1}{4} \bm c_1 
 + \frac{1}{4} \bm c_2 ,
\nonumber 
\end{eqnarray}
%===================================================================
The vectors 
$\pm \bm v(e_i)$ and 
$\pm \bm v(f_i)$ 
constitute the building block of the $K_4$ crystal, e.g.,
$\bm v(e_1)$ is the vector connecting the $A$ and $B$ sites 
  in Fig.\ \ref{fig1}(b).

One possible choice of 
$\bm c_i$ as
$\bm c_1 = (+1,-1,-1)$, 
$\bm c_2 = (-1,+1,-1)$, and
$\bm c_3 = (-1,-1,+1)$.
The realized $K_4$ crystal is shown in Fig.\ \ref{fig1}(b).
The numbers in the parentheses represent the fractional 
Cartesian coordinates
  in the cubic conventional unit cell, where we note that 
the lattice constant becomes  $a=2$ by the definition
of the building block vectors [Eq.\ (\ref{eq:vectors})].
From the crystallographic point of view,
the space group of the $K_4$ crystal is $I4_132$ (No.\ 214) and
the primitive vectors are chosen as 
$\bm t_1=(-\frac{1}{2},\frac{1}{2},\frac{1}{2})$,
$\bm t_2=(\frac{1}{2},-\frac{1}{2},\frac{1}{2})$, and 
$\bm t_3=(\frac{1}{2},\frac{1}{2},-\frac{1}{2})$, in the unit $a=2$.
There are four sites in the primitive cell and
these four sites are specified as $A$, $B$, $C$, and $D$, 
in Fig.\ \ref{fig1}.
The coordination number is three, 
as in the 2D graphene.

The $K_4$ crystal has the remarkable mathematical property, called 
the \textit{strongly isotropic} property 
\cite{Sunada:2007tb,Sunada:2012ve}, indicating
the property that preserves the crystal net after 
any permutation of bonds with common end point.
For example, by focusing of the $A$ site at
$(\frac{1}{8},\frac{1}{8},\frac{1}{8})$
in Fig.\ \ref{fig1}(b),
we can keep the crystal net even if we exchange 
  the bonds $\bm v(e_2)$ and $\bm v(e_3)$
 while the $\bm v(e_1)$ bond is fixed.
This congruent transformation can be realized by combination of  
the rotation about the twofold rotation axis $C_{2f}$ \cite{Bradley_Group}
and the subsequent translation
$\bm r\to \bm r+\bm t$ 
where $\bm t=(\frac{1}{4},\frac{1}{4}, \frac{1}{4})$.
In the usual crystallographic notation,
this transformation is represented by
$\{C_{2f}|\frac{1}{2}\frac{1}{2}\frac{1}{2} \}$,
which is 
nothing but one element of the space group $I4_132$.
Here, we note that the translation vector $\frac{1}{2}\frac{1}{2}\frac{1}{2}$
is given in the unit of the primitive vectors, i.e.,
 $\bm t = \frac{1}{2}\frac{1}{2}\frac{1}{2}
=\frac{1}{2}\bm t_1
 +\frac{1}{2}\bm t_2
 +\frac{1}{2}\bm t_3=(\frac{1}{4},\frac{1}{4}, \frac{1}{4})$.
It is known that the honeycomb and the diamond crystals
 have this strongly isotropic property
\cite{Sunada:2007tb,Sunada:2012ve}.
The honeycomb is the only crystal having the strongly isotropic
property in 2D,
and there are only two strongly isotropic crystals in 3D: 
one is the diamond crystal and the other is 
the $K_4$ crystal.
In this sense, 
the diamond crystal and $K_4$ crystal are the most \textit{beautiful}
crystals in 3D and 
the $K_4$ crystal can be called the \textit{diamond twin}
\cite{Sunada:2012ve}.
It has been emphasized \cite{Sunada:2007tb,Sunada:2012ve}
that  the $K_4$ crystal has chirality as can be seen from the existence of
 the $4_1$ screw axis, in contrast to the diamond crystal.
Thus the effect of the SOC on the $K_4$ crystal is different from that 
  on the diamond crystal
due to the lack of the I symmetry.
We argue the effect of the SOC in Sec.\ \ref{sec:SOC}.

%====================================================================
\begin{figure}[t]
\includegraphics[width=8.5cm,bb=24 15 486 776]{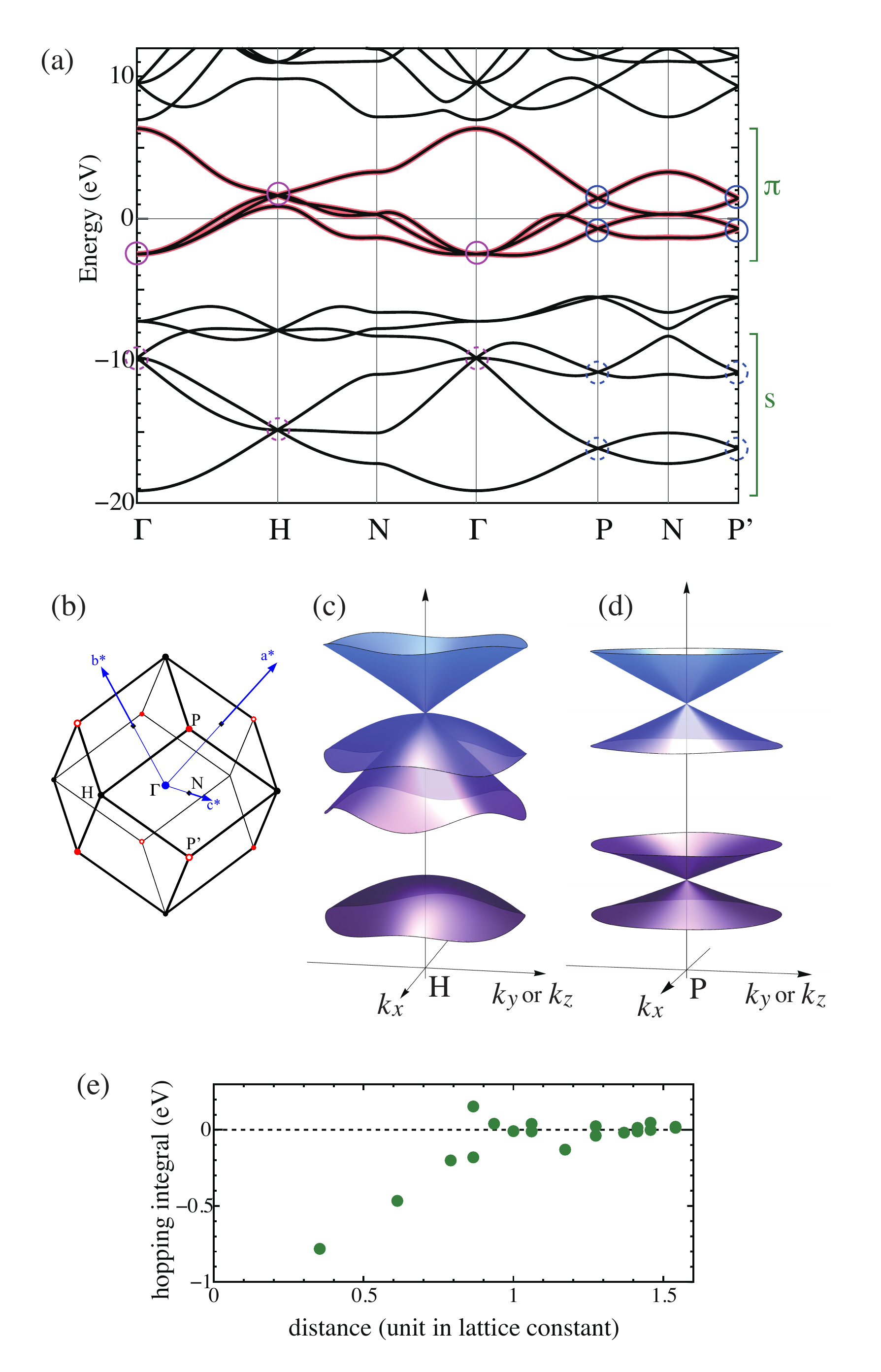}
 \caption{
(a) Band structure for the $K_4$ carbon obtained from the
 first-principles calculation.
The band structure by 
the tight-binding model is 
 shown by the bold (red) curves.
(b) The Brillouin zone for bcc.
The band structures near the $H$ (c) and $P$ (d) points.
(e) The evaluated hopping integrals as a function of the intersite distance.
}
\label{fig2}
\end{figure}
%======================================================================

\section{$\bm K_4$ carbon}\label{sec:K4carbon}

In this section, we construct 
the tight-binding model for the 
 $K_4$ carbon in terms of the first-principles calculation.

\subsection{Band structure of the $K_4$ carbon}

The stability of the $K_4$ carbon 
 has been discussed theoretically 
and the metallic behavior was predicted
\cite{Rignanese:2008ii,Itoh:2009gs}.
The optimized bond distance is $~ 1.4-1.5$ \AA, which is comparable to 
that in  diamond and graphite.
Figure \ref{fig2}(a) shows 
the band structure obtained by the 
first-principles density-functional-theory (DFT) calculation based on
 generalized gradient approximation with the use of the \texttt{WIEN2k} code
\cite{wien2k}.
The lattice constant is set as $a=4.063$ \AA, according to Ref.\
 \cite{Rignanese:2008ii},
where the bond distance for nearest-neighbor sites 
is $d\approx 1.44$ \AA.
The Brillouin zone is shown in Fig.\ \ref{fig2}(b).
The band structure well reproduce the ones reported in 
 Refs.\ \onlinecite{Rignanese:2008ii} and \onlinecite{Itoh:2009gs}.
The four conduction bands in the energy range 
$-3.5\mbox{ eV} < E_k < 6.5 \mbox{ eV}$
are constructed by the carbon 
$p$ orbitals that stand perpendicular to the plane formed by the
 nearest-neighbor carbon atoms, i.e., by the ``$\pi$'' orbitals.
The valence bands in the energy range 
$-20\mbox{ eV} < E_k \lesssim -8 \mbox{ eV}$
 are mainly formed by the carbon $s$ orbitals. 
The bands at $-8\mbox{ eV} \lesssim E_k < -5 \mbox{ eV}$
are constructed by both the $s$ orbitals and the $p$ orbitals elongated
perpendicular to the $\pi$ orbital, i.e., 
by  the ``$\sigma$'' hybridized orbitals.
We observe two kinds of the nontrivial degenerate points in the conduction bands.
First, 
at the $H$ ($\Gamma$) point,
  the bands are triply degenerate with $E=1.63$ eV ($-2.48$ eV).
The band structure near the $H$ point is 
  explicitly shown in Fig.\ \ref{fig2}(c).
We find that  
 the bands near the $H$ point 
     exhibit the linear $\bm k$ dependencies 
  except for the middle band.
Due to the 3D isotropic structure, the band structures are isotropic 
with the axes $k_x$, $k_y$, and $k_z$.
Secondly, 
a pair of the degenerate two bands can be observed 
 at the $P$ point, as shown in Fig.\ \ref{fig2}(d).
We also note that the triply and doubly degenerate points can also be observed 
in the valence bands.

\subsection{Tight-binding model  of the $K_4$ carbon}

In order to analyze the band structure in more detail, 
 we construct the tight-binding model based on the maximally localized
 Wannier functions \cite{Kunes20101888}, by targeting the four
 conduction bands
 in the energy range $-3.5\mbox{ eV} < E_k < 6.5 \mbox{eV}$.
The Wannier functions contributing these four bands are well described by 
the  $\pi$ orbitals.
The standing directions of the $\pi$ orbitals for the $A$--$D$ sites
can be described by the normal vectors 
$\bm n(A) = (1,1,1)$,
$\bm n(B) = (-1,1,1)$,
$\bm n(C) = (1,-1,1)$, and
$\bm n(D) = (1,1,-1)$.
These four $\pi$ orbitals in the primitive unit cell
construct  four conduction bands.
The tight-binding hopping integrals evaluated on the basis of the 
  Wannier functions \cite{Kunes20101888} are shown in Fig.\ \ref{fig2}(e).
The hopping parameter for the nearest neighbor sites is  
  given by $t_1=-0.782$ eV.
This parameter can be contrasted to the one in the graphene 
\cite{Reich:2002kd,Jung:2013ix}:
recent evaluation of the hopping parameters for 
2D graphene indicates 
$(pp\pi) \approx -3.0$ eV \cite{Jung:2013ix}.
The hopping amplitude depends on the angle between the $\pi$ orbitals 
  of the neighboring carbon atoms.
In terms of the Slater-Koster parametrization \cite{Slater:1954ui}, 
the nearest-neighbor hopping integral for the $K_4$ carbon 
  is given by $t_1=(pp\pi)/3$.
Thus the parameter of 
$(pp\pi)$
for the $K_4$ carbon is consistent
  with that for the graphene.  
In addition, we observe that the amplitudes of the 
long-distance hoppings are relatively large.
This fact is also consistent with the results in the graphene
\cite{Jung:2013ix}.

The band structures obtained from the tight-binding approximation
  are shown by 
bold (red) curves in Fig.\ \ref{fig2}(a).
We find the tight-binding model based only on the $\pi$ orbital
perfectly reproduces the DFT results of the conduction bands.
There is presumably small but nonzero hopping between 
the $\pi$ and $s$ orbitals with different sites.
Such an effect would be included effectively and, as a result, 
 the relatively large long-distance hopping parameters
are obtained.
This would be one reason why the four $\pi$-orbital description works well.

%====================================================================
\begin{figure*}[t]
\includegraphics[width=17cm,bb=22 16 730 258]{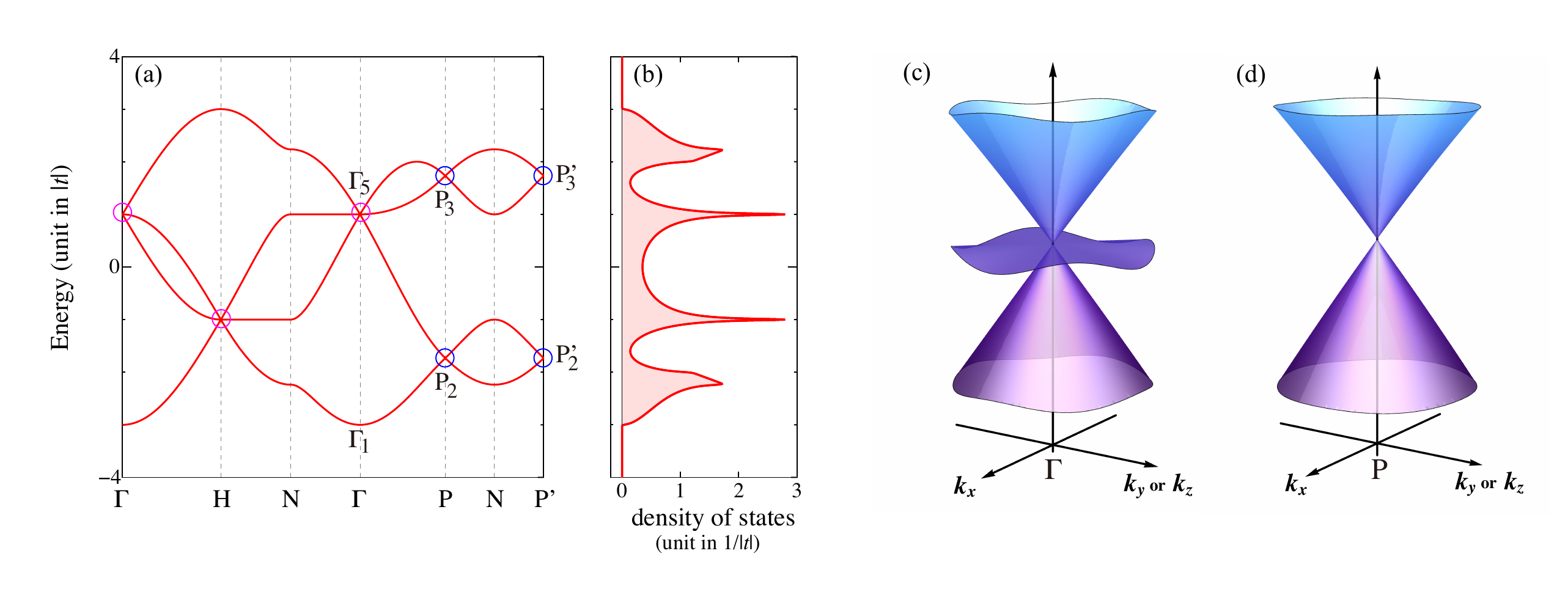}
 \caption{
(a)
The band structures of the tight-binding model for the $K_4$ crystal.
Only the nearest-neighbor hoppings are taken into account.
The Dirac points are indicated by circles.
The triple degeneracy at the $\Gamma_5$ point 
is described by the three-dimensional real $T_2$
 irreducible representation, and the double degeneracies at 
the $P_2$ and $P_3$ points are described by the two-dimensional 
 complex ${}^1F_2$ and ${}^1F_3$ irreducible representations.
The $P_2$ and $P'_2$  ($P_3$ and $P'_3$) are the conjugate pair 
  of the complex representation.
(b) The density of states.
The case for half filling $n=1$ is shown where $n$ is the filling factor.
The Fermi energy coincides with that of the
 $S=1$ Dirac point if $n=2/3$ or $4/3$. 
(c) The $S=1$ Dirac cone at the $\Gamma$ and $H$ points.
(d) The $S=1/2$ Dirac cone at the $P$ points.
}
\label{fig3}
\end{figure*}
%======================================================================

\section{Dirac points in the $\bm K_4$ crystal}\label{sec:Dirac}

In the section, we analyze the $\Gamma$ and $P$ points 
on the basis of the tight-binding model and 
show that the degenerate points at $\Gamma$ and $P$ points 
  are described by the pseudospin $S=1$ and $S=1/2$ Dirac cones,
respectively.
A generalization of the 
  Dirac cone structure to posses 
 pseudospin $S>\frac{1}{2}$
has been discussed in the literature
\cite{Watanabe:2011fd,Orlita:2014br,Malcolm:2014fn}.
In the conventional $S=1/2$ Dirac cone, \textit{two} bands 
  exhibit linear dependence in the momentum and touch at a single point.
In contrast, in the $S=1$ case, \textit{three} bands become degenerate and 
touch at a single point, where 
  anomalous physical behavior can be expected \cite{Malcolm:2014fn}.
Recently a possible system to emerge the $S=1$ Dirac cone 
  has been proposed  in terms of the first-principles calculation
 \cite{Giovannetti:2015by}.
However, 
the explicit tight-binding model that
exhibits the $S=1$ Dirac cone is not obtained, and
 then 
the physical properties have not been clarified yet.

In order to simplify the discussions, 
  we focus on the nearest-neighbor hopping only.
This situation is indeed relevant to the 
  recently discovered $K_4$ crystal
\cite{Mizuno:2015ip}, as will be discussed later.
Since there are four sites in the primitive unit cell, 
  the Hamiltonian can be described as the 
  $4\times 4$ matrix in the orbital basis
$\biglb(\varphi_{A}(\bm k),\varphi_{B}(\bm k),
   \varphi_{C}(\bm k),\varphi_{D}(\bm k)\bigrb)$,
Here we consider  the $s$-orbital bands.
The tight-binding Hamiltonian is explicitly given by
%===================================================================
\begin{equation}
 H_{\bm k}
=
-
\left(
\begin{array}{cccc}
0 
&  e^{-\frac{i}{2}\bm k \cdot \bm v(e_1)}
&  e^{-\frac{i}{2}\bm k \cdot \bm v(e_2)} 
&  e^{-\frac{i}{2}\bm k \cdot \bm v(e_3)} 
\\
   e^{\frac{i}{2}\bm k \cdot \bm v(e_1)}
&  0
&  e^{-\frac{i}{2}\bm k\cdot \bm v(f_3)}
&  e^{\frac{i}{2}\bm k\cdot \bm v(f_2)}
\\
  e^{\frac{i}{2}\bm k \cdot \bm v(e_2)} 
& e^{\frac{i}{2}\bm k\cdot \bm v(f_3)}
&  0
& e^{-\frac{i}{2}\bm k\cdot \bm v(f_1)}
\\
  e^{\frac{i}{2}\bm k \cdot \bm v(e_3)} 
& e^{-\frac{i}{2}\bm k\cdot \bm v(f_2)}
& e^{\frac{i}{2}\bm k\cdot \bm v(f_1)}
& 0
\end{array}
\right).
\label{eq:H_tb}
\end{equation}
%===================================================================
Owing to the property of completeness of $K_4$, the off-diagonal 
  components of the matrix become dense.
The nearest-neighbor hopping parameter is set to $t=-1$.
The building block vectors [Eq.\ (\ref{eq:vectors})] are explicitly
given by
$\bm v(e_1) = \frac{1}{2} (0,-1,1)$,
$\bm v(e_2) = \frac{1}{2} (1,0,-1)$,
$\bm v(e_3) = \frac{1}{2} (-1,1,0)$,
$\bm v(f_1) = \frac{1}{2} (0,-1,-1)$,
$\bm v(f_2) = \frac{1}{2} (-1,0,-1)$, and
$\bm v(f_3) = \frac{1}{2} (-1,-1,0)$.
In the $s$-orbital case, the sign of the hopping is common.
If we consider the $\pi$-orbital case,
the sign of the hopping integral can be altered
depending on bonding or anti-bonding character of the $\pi$ overlapping.
The signs of the hoppings are determined by 
those of the inner product of vectors, i.e.,
$\bm n(B)\cdot \bm n(C) = - 1 $,
$\bm n(B)\cdot \bm n(D) = - 1 $, 
$\bm n(C)\cdot \bm n(D) = - 1 $, and become positive otherwise.
Thus in the case of $\pi$-orbital case
the extra prefactor $-1$ should be added 
 for the 
(2,3), (2,4), (3,2), (3,4),
 (4,2), (4,3) 
matrix components.

The band structures obtained from Eq.\ (\ref{eq:H_tb}) 
 are shown in Fig.\ \ref{fig3}(a).
The degenerate dispersion relations near 
 the $\Gamma$ and $H$ points ($P$ and $P'$ points)
can be described by the $S=1$ ($S=1/2$) Dirac cone
 as shown shortly.
The $S=1$ Dirac point is at the $\Gamma_5$ point
  with the energy $E=+1$ where the bands are triply degenerate.
The triple degeneracy at the $\Gamma_5$ point is 
  described by the three-dimensional real $T_2$ irreducible representation
\cite{Bradley_Group}.
The same profile can be seen at the $H$ point.
The $S=1/2$ Dirac points are at $P_2$ and $P_3$ points 
  where the bands are doubly degenerate at the energy $E=\pm \sqrt{3}$.
The double degeneracies at $P_2$ and $P_3$ are described by the 
two-dimensional complex ${}^1F_2$ and ${}^1F_3$ irreducible representations
\cite{Bradley_Group}.
Due to the presence of the flat bands near the  $S=1$ Dirac points at
 $\Gamma$ and $H$, the critical enhancement can be observed 
 at $E=\pm 1$  in density of states (DOS), as shown in 
Fig.\ \ref{fig3}(c).
In contrast, sufficient suppressions can be seen at $E=\pm \sqrt{3}$,
  reflecting the presence of the $S=1/2$ Dirac cones.
The same profile of the DOS can be seen in the 3D hyperkagom\'e  crystal 
except for the van Hove singularity owing to flat bands 
\cite{Udagawa:2009gv}.
Note that the DOS does not vanish precisely at $E=\pm \sqrt{3}$, 
  since another band across this energy at different position 
  in $\bm k$.

%====================================================================
\begin{figure}[t]
\includegraphics[width=8cm,bb=68 220 765 426]{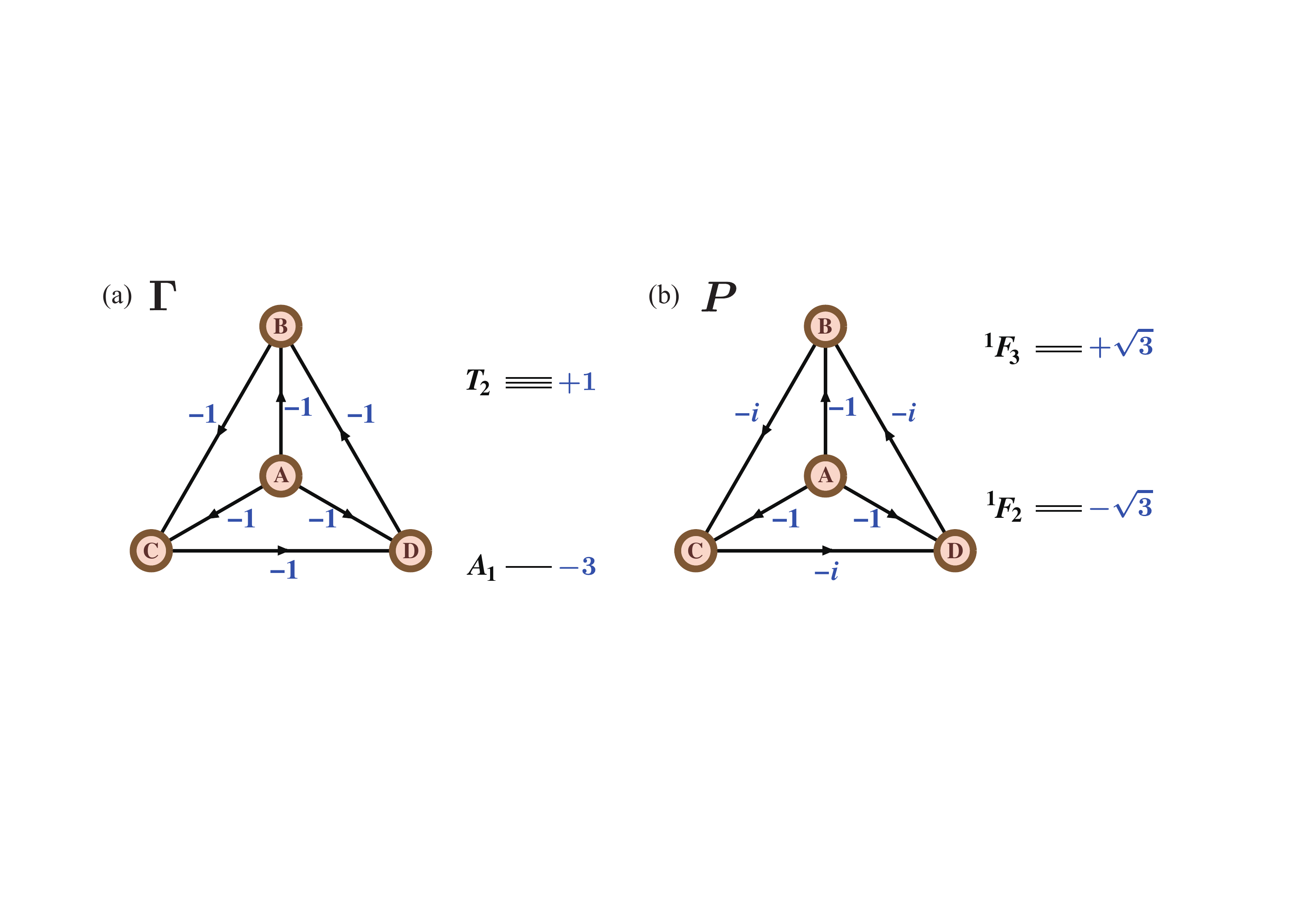}
 \caption{
Structure of the Hamiltonian matrix
and the corresponding energy diagrams 
 at the $\Gamma$ point (a) and  the $P$ point (b).
 }
\label{fig4}
\end{figure}
%======================================================================

\subsection{$\bm{S=1}$ Dirac cone at the $\bm \Gamma$ point}

Here we derive 
the effective Hamiltonian near the $\Gamma$  point and show 
explicitly that
  its character is described by pseudospin $S=1$ Dirac cone.
By setting $\bm k=(0,0,0)$ in Eq.\ (\ref{eq:H_tb}),
all the off-diagonal matrix elements become $-1$,
and then the eigenvalues are $-3$ 
(no degeneracy)   and $+1$ (triple degeneracy)
[see Fig.\ \ref{fig4}(a)].
The energy separation at the $\Gamma$ point can be
recognized by regarding 
the $K_4$ graph [Fig.\ \ref{fig1}(a)] as
a regular tetrahedron.
The eigenfunctions of the regular tetrahedron 
can be classified according to the 
representation of the point group $\mathrm{T_d}$ and 
  are composed of the $A_1$ representation and the $T_2$ representation.
In the tight-binding picture, the $A_1$ representation has the energy $-3$
 and the $T_2$ representation have $+1$.
One of the possible choices of the unitary matrix 
for diagonalizing $H_{\bm k}$ on the  $\Gamma$  point
is given by
%===================================================================
\begin{eqnarray}
 U_\Gamma
=
\frac{1}{2}
\left(
\begin{array}{cccc}
   1
& e^{-i\frac{\pi}{4}}
& e^{+i\frac{\pi}{2}}
& e^{+i\frac{\pi}{4}}
\\
   1
&  e^{-i\frac{3\pi}{4}}
&  e^{-i\frac{\pi}{2}}
&  e^{+i\frac{3\pi}{4}}
\\
   1
&  e^{+i\frac{\pi}{4}} 
&  e^{-i\frac{\pi}{2}}
&  e^{-i\frac{\pi}{4}}
\\
   1 
&  e^{+i\frac{3\pi}{4}}
&  e^{+i\frac{\pi}{2}}
&  e^{-i\frac{3\pi}{4}}
\end{array}
\right).
\label{eq:Ugamma}
\end{eqnarray}
%===================================================================
The first column corresponds to  the $A_1$ representation and 
the remaining three columns to the $T_2$ representation of the 
point group $\mathrm{T_d}$.
By applying this unitary matrix to the Hamiltonian (\ref{eq:H_tb}), 
  and by expanding the momentum up to $O(\bm k)$, we obtain
%===================================================================
\begin{eqnarray}
U_\Gamma^\dagger H_{\bm k} U_\Gamma
\!\!\! &=& \!\!\!
\hat E_\Gamma 
+ \frac{1}{2}
\left(
\begin{array}{cccc}
0
&  0
&  0
&  0
\\
  0
&  k_z
&  \frac{1}{\sqrt{2}} k_-
& 0
\\
  0
& \frac{1}{\sqrt{2}} k_+
&  0
& \frac{1}{\sqrt{2}} k_-
\\
0 
& 0
& \frac{1}{\sqrt{2}} k_+
& -k_z
\end{array}
\right)
+O(k^2),
\nonumber \\
\label{eq:HGamma}
\end{eqnarray}
%===================================================================
where $k_\pm \equiv (k_x\pm ik_y)$,
and 
 $\hat E_\Gamma =
\mathrm{diag}(-3,1,1,1)$
  is the set of the energy eigenvalues on the $\Gamma$ point.
If we focus on the second, third, and fourth rows and columns 
  in the second term in Eq.\ (\ref{eq:HGamma}),
the effective Hamiltonian is given by
  the $3\times 3$ matrix:
%===================================================================
\begin{equation}
H_{\mathrm{eff}} = \frac{1}{2} \bm k \cdot \bm{S} ,
\end{equation}
%===================================================================
where 
$\bm S=(S_x,S_y,S_z)$ is the spin-$1$ matrix:
%===================================================================
\begin{eqnarray}
&
S_x =
\left(
\begin{array}{ccc}
0  & \frac{1}{\sqrt{2}} & 0
\\
\frac{1}{\sqrt{2}}  & 0 & \frac{1}{\sqrt{2}} 
\\
0  & \frac{1}{\sqrt{2}}  & 0
\end{array}
\right), \quad
S_y =  
\left(
\begin{array}{ccc}
0  & \frac{-i}{\sqrt{2}}  & 0
\\
\frac{i}{\sqrt{2}}  & 0 & \frac{-i}{\sqrt{2}} 
\\
0  & \frac{i}{\sqrt{2}}  & 0
\end{array}
\right),&
\nonumber \\
&S_z =
\left(
\begin{array}{ccc}
1  & 0 & 0
\\
0  & 0 & 0
\\
0  & 0 & -1
\end{array}
\right).&
\end{eqnarray}
%===================================================================
Thus the electronic structure near the $\Gamma$ point 
is described by the pseudospin $S=1$ Dirac cone.
The energy dispersions are given by
%===================================================================
\begin{eqnarray}
E_{\bm k}^+ &=& 1 + \frac{1}{2} \sqrt{k_x^2 + k_y^2 + k_z^2 } +  O(k^2),
\nonumber
\\
E_{\bm k}^0 &=& 1 + O(k^2),
\\
E_{\bm k}^- &=& 1 - \frac{1}{2} \sqrt{k_x^2 + k_y^2 + k_z^2 } + O(k^2).
\nonumber
\end{eqnarray}
%===================================================================
The dispersion relations near 
 the $H$ point are described in a similar manner.
The $S=1$ Dirac cone structure near the $\Gamma_5$ point
is shown in Fig.\ \ref{fig3}(c).

\subsection{$\bm S=\frac{1}{2}$ Dirac cone at the $\bm P$ point}

Next, we focus on the dispersion relations near the $P$ point.
By setting $\bm k=\bm P \equiv (\pi,\pi,\pi)$
in Eq.\ (\ref{eq:H_tb}),
  some matrix elements become imaginary
[see Fig.\ \ref{fig4}(b)] 
and the eigenvalues are 
$\pm \sqrt{3}$
with double degeneracy.
We find that the conjugate pair of 
the eigenfunctions 
$(\varphi_A,\varphi_B,\varphi_C,\varphi_D)
=(0,1,\omega,\omega^2)$ and 
 $(0,1,\omega^2,\omega)$, where  $\omega\equiv \exp(i2\pi/3)$,
have different energies.
The degenerate pairs can be generated by the $C_2$ transformation of the
regular tetrahedron, e.g., 
 $(A,B,C,D)\to(C,iD,A,-iB)$. 
Then the full eigenfunctions are
  $(\varphi_A,\varphi_B,\varphi_C,\varphi_D)=
   (0,1,\omega,\omega^2)$ and $(  \omega, i\omega^2,0,-i)$ 
 for $E_{\bm k}=+\sqrt{3}$, and
  $(\varphi_A,\varphi_B,\varphi_C,\varphi_D)=
   (0,1,\omega^2,\omega)$ and  $(  \omega^2, i\omega,0,-i)$ 
 for $E_{\bm k}=-\sqrt{3}$.
From the Gram-Schmidt orthogonalization procedure,
one of the choices of 
the unitary matrix  for diagonalizing $H_{\bm k}$ on the $P$
point
is given by
%===================================================================
\begin{eqnarray}
 U_P
&=&
\frac{1}{\sqrt{6}}
\left(
\begin{array}{cccc}
  \sqrt{3}
&  0
&  -\sqrt{3}
&  0
\\
    1
&   \sqrt{2}
&   1
&  \sqrt{2} \omega^2
\\
    1
&   \sqrt{2}\omega^2
&   1
&   \sqrt{2}
\\
 1
&  \sqrt{2}\omega
&  1
&  \sqrt{2}\omega
\end{array}
\right).
\end{eqnarray}
%===================================================================
The first and second (third and fourth) columns represent the 
  eigenvectors for the eigenvalue $-\sqrt{3} (+\sqrt{3})$.
In order to analyze the dispersion relation near this point,
we apply the  unitary transformation  to the Hamiltonian
(\ref{eq:H_tb}).
By expanding it up to the first order in the momentum $\bm k$, we find
\begin{widetext}
%===================================================================
\begin{equation}
U_P^\dagger H_{\bm P+\bm k} U_P
=
\hat E_P
+
\frac{1}{6}
\left(
\begin{array}{cccc}
k_x+k_y+k_z
& 
\sqrt{2}
(k_x +  \omega^2 k_y + \omega k_z)
&  
k_x+k_y+k_z
&  
-
\frac{1}{\sqrt{2}}
(\omega^2 k_x
+ k_y
+ \omega k_z
)
\\ %------
&  
-(k_x+k_y+k_z)
&  
-
\frac{1}{\sqrt{2}}
( k_x
+ \omega k_y
+ \omega^2  k_z
)
&  
-
( \omega^2 k_x
+ \omega  k_y
+ k_z
)
\\
&  
&  
k_x+k_y+k_z
&  
\sqrt{2}
( \omega^2 k_x 
+  k_y
+ \omega k_z )
\\
& 
& 
& 
-(k_x+k_y+k_z)
\end{array}
\right)
+O(k^2),
\label{eq:UHU}
\end{equation}
%===================================================================
\end{widetext}
where $\hat E_P =
\mathrm{diag}(-\sqrt{3},-\sqrt{3},+\sqrt{3},+\sqrt{3})$
  is the set of the energy eigenvalues on the $P$ point.
Here the quantity $\bm k$ indicates the momentum centered at the $P$ point.
From the perturbative arguments  up to $O(k_i)$, 
 we can neglect
  the off-diagonal matrix elements connecting 
  the states with different eigenvalues $\pm \sqrt{3}$,
since the contributions of the dropped terms are of the order of $k^2$.
Thus
 the Hamiltonian (\ref{eq:UHU}) can be divided into two 
$2\times 2$ Hamiltonians.
The effective Hamiltonian  representing  lower two bands
  is given by
%===================================================================
\begin{equation}
  H_{\bm k}^\mathrm{eff}
=
\frac{1}{6}
\left(
\begin{array}{cc}
k_x+k_y+k_z
& 
\sqrt{2} (k_x + \omega^2 k_y + \omega k_z)
\\
\sqrt{2} (k_x + \omega k_y + \omega^2 k_z)
&  
-k_x-k_y-k_z
\end{array}
\right),
\end{equation}
%===================================================================
From this Hamiltonian, we immediately find that
the energy dispersion is given by
%===================================================================
\begin{eqnarray}
E_{\bm k,\pm}^\mathrm{eff}
=
\pm \frac{1}{2\sqrt{3}} \sqrt{k_x^2+k_y^2+k_z^2}
+ O(k^2),
\end{eqnarray}
%===================================================================
which represents the 3D $S=1/2$ Dirac cone.
The dispersion relation near the $P$ point
is shown in Fig.\ \ref{fig3}(d).
In the case of the $P'$ point, the Dirac cone has opposite chirality,
 as in the case of 2D graphene.

In the above analysis, we focused only on the
nearest-neighbor hopping.
By taking into account the long-distance hopping parameters  
  shown in Fig.\ \ref{fig2}(e), the conduction $\pi$ band dispersions 
can be reproduced 
[the bold (red) curves in Fig.\ \ref{fig2}(a)].
We find that
the Dirac cone structures are robust against the
long-distance hoppings.
Here we note that, in the case of the $\pi$ orbital, the
 structures at $\Gamma$ and $H$ points are interchanged.
Since the long-distance hopping parameters are relatively large, 
 the  band structures  are strongly modified. 
Especially, the energy level with the $A_1$ representation 
($\Gamma_1$ point) becomes
 higher than that of the $S=1/2$ Dirac point at the $P_2$ point
 in the case of the carbon system.
However,
in the case of the recently discovered $K_4$ crystal
\cite{Mizuno:2015ip},
the long-distance hopping can be 
small and this simple treatment based only on the 
nearest-neighbor hopping
can be justified,
as will be discussed later.

%====================================================================
\begin{figure}[t]
\includegraphics[width=8.5cm,bb=23 23 967 492]{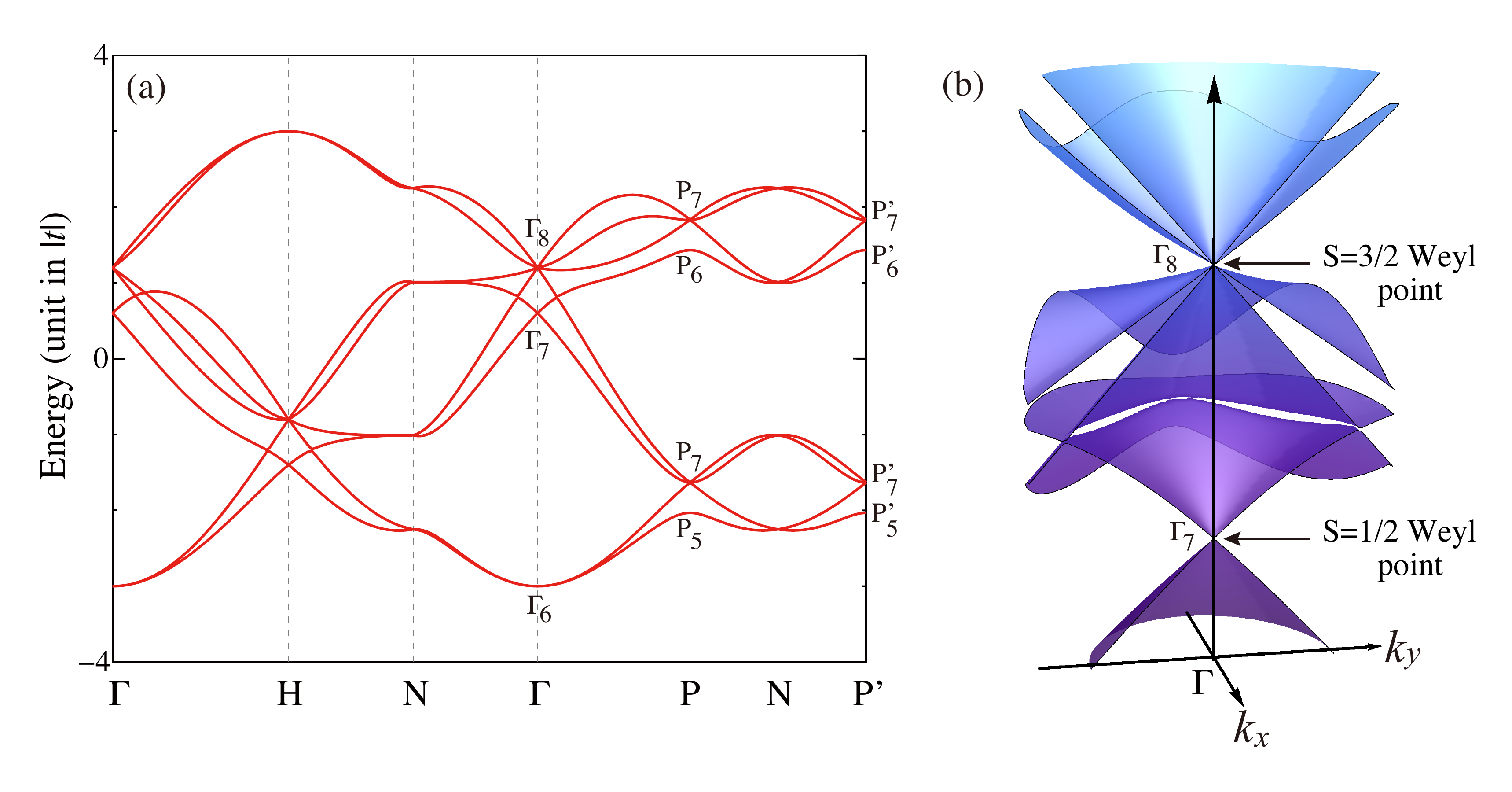}
 \caption{
(a) Band structure of the tight-binding model for the $K_4$ crystal
with the spin-orbit interaction $\lambda_\mathrm{SO}=0.05$.
In the double-valued representation,
the $\Gamma_6$ and $\Gamma_7$  points are described by 
 the two-dimensional $\bar E_1$ and $\bar E_2$ irreducible 
 representation, respectively, and 
  the $\Gamma_8$ point is by the four-dimensional $\bar F$ representation.
The $P_5$ and $P_6$ points are the conjugate pair of the one-dimensional 
  ${}^1 \bar E$ and ${}^2 \bar E$ representation.
The triple degeneracy at the $P_7$ points
is described by the three-dimensional $\bar T$ representation.
(b) The dispersions near the $\Gamma_7$ and $\Gamma_8$ point are shown
 with fixed $k_z=0$.
In this figure, only the region $k_x<0$ is shown.
 }
\label{fig5}
\end{figure}
%======================================================================

\section{Effect of the spin-orbit coupling}\label{sec:SOC}

In this section, we analyze the effect of the SOC
 in the $K_4$ crystal.
The effect of the SOC is not common even within 
the strongly isotropic crystals.
The Dirac points in the 2D graphene are not robust against 
  the SOC \cite{Kane:2005hl}; on the other hand,
the Dirac points can emerge in 3D diamond as a consequence of the SOC
 \cite{Young:2012kz}.
In contrast to the 2D honeycomb  and 3D diamond crystals,
  the $K_4$ crystal does not hold the inversion symmetry
\cite{Sunada:2007tb}.
Thus the degeneracy of band at general points is lifted 
  due to the SOC,
and the modification of the band structure near the Dirac points  
 shows unique properties.

As has been discussed in Refs.\ 
\cite{Kane:2005hl,Fu:2007io,Young:2012kz},
the intrinsic SOC can be expressed in terms of
spin-dependent next-nearest-neighbor hopping. 
The explicit Hamiltonian of the SOC is given by
%===================================================================
\begin{eqnarray}
H_\mathrm{SOC}
=
i \frac{8\lambda_\mathrm{SO}}{a^2} 
\sum_{\langle\langle ij\rangle\rangle}
c_i^\dagger \, \bm s \cdot (\bm v_{ij}^1 \times \bm v_{ij}^2) \, c_j ,
\label{eq:SOC}
\end{eqnarray}
%===================================================================
where $\langle\langle ij\rangle\rangle$ represents the summation 
of the sites over the next-nearest-neighbor pairs, and 
$\bm v_{ij}^{1,2}$ are the nearest-neighbor bond vectors traversed 
between sites $i$ and $j$.
The band structure of the tight-binding model (\ref{eq:H_tb}) in the 
 presence of the SOC term (\ref{eq:SOC}) is shown in Fig.\ \ref{fig5}(a).
In the general points of $\bm k$, the degeneracy of the bands is
lifted due to the SOC since the $K_4$ crystal does not have the
inversion symmetry.

The band structures near the $\Gamma$ point 
show unique properties. 
When the SOC is introduced,
the sixfold degeneracy at the $\Gamma_5$ point
  splits into two energy levels of
$E_{\Gamma_7} = +1 - 8 \lambda_\mathrm{SO}$
with the two-dimensional 
 $\Gamma_7$ ($\bar E_2$)
representation 
 and 
$E_{\Gamma_8}=+1 + 4 \lambda_\mathrm{SO}$ with 
the four-dimensional $\Gamma_8$ ($\bar F$)  representation, 
whereas the
$\Gamma_1$ point turns into the two-dimensional $\Gamma_6$ ($\bar E_1$)
with the energy $E_{\mathrm k}=-3$. 
The effective Hamiltonian near the $\Gamma$ point
can be obtained 
by applying the transformation given by Eq.\ (\ref{eq:Ugamma}) to 
 the SOC term [Eq.\ (\ref{eq:SOC})].
Especially near the $S=1$ Dirac point ($\Gamma_5$), 
the SOC induces a perturbation 
in the form of a $6\times 6$ matrix:
%===================================================================
\begin{eqnarray}
H_\mathrm{SOC}^\mathrm{eff}
=
4\lambda_\mathrm{SO}\, \bm{S} \cdot \bm{\sigma},
\end{eqnarray}
%===================================================================
where $\bm \sigma=(\sigma_x,\sigma_y,\sigma_z)$ is the 
 $2\times 2$ Pauli matrix representing the electron's spin.
The energy dispersion near the $\Gamma_7$ and $\Gamma_8$ points 
is shown in Fig.\ \ref{fig5}(b).
Near the fourfold degenerate $\Gamma_8$ point, 
the band structure exhibits the linear $\bm k$ dependencies.
The effective Hamiltonian near the $\Gamma_8$ point can be derived 
 by considering the situation $|\bm{k}|\ll |\lambda_\mathrm{SO}|$ 
and is described by a $4\times 4$ matrix.
By applying an appropriate unitary transformation 
  for diagonalizing the SOC term, the effective Hamiltonian is given by
 $H^\mathrm{eff}_\mathrm{\Gamma_8}= 
1+4\lambda_{\mathrm{SO}}
+ \frac{1}{3}\bm k\cdot \bm{J}$, where
$\bm{J}=(J_x,J_y,J_z)$ is the $4\times 4$ spin-$3/2$  matrix.
Therefore the dispersions near the $\Gamma_8$ point
 are described by the $S=3/2$ Weyl dispersions, where
the dispersions take forms
$1+4\lambda_\mathrm{SO}\pm \frac{1}{2} |\bm{k}|$ and
$1+4\lambda_\mathrm{SO}\pm \frac{1}{6} |\bm{k}|$.
Incidentally, we can observe $S=1/2$ Weyl dispersions
  around the $\Gamma_7$ point,  where
the energy dispersions are given by
$1 -8 \lambda_\mathrm{SO}\pm \frac{1}{3} |\bm k|$.
Similar structure can be observed at the $H$ point.

At the $P$ point, we also observe the unique properties.
The energy splitting on the $P$ points are given by
$\pm \sqrt{3}-6\lambda_\mathrm{SO}$ (unique) 
at the  $P_5$ or $P_6$ point
and
$\pm \sqrt{3}+2\lambda_\mathrm{SO}$ (triply degenerate)
at the $P_7$ point.
The $P_5$ and $P_6$ points are  described by 
the conjugate pair of  the one-dimensional complex $^1\bar E$ and $^2\bar E$ 
representations.
On the other hand,
 the $P_7$ point is described by the 
  $\bar T$ representation
and  would be described by the $S=1$ Weyl point.
Incidentally, we observe several contact points with 
 accidental degeneracy 
 at general $\bm k$ points, e.g., along the $\Gamma$-$P$ and 
$\Gamma$-$N$ lines.
Analyses of physical quantities on this system are 
desired for future work.

\section{Summary}\label{sec:summary}

In summary, we have examined the energy dispersion of the $K_4$ crystal 
  in detail. 
The tight-binding model has been derived explicitly where we show
the emergence of  
 the pseudospin $S=1$ and $S=1/2$ Dirac cones.
We have also analyzed the effect of SOC to examine how 
 the degeneracies at the Dirac points are lifted.
In contrast to the other strongly isotropic 
honeycomb and diamond crystals, the $K_4$ crystal lacks the inversion symmetry, 
and the lowering of the symmetry is quite peculiar.
We found that,
by including the SOC, 
the $S=1$ Dirac point split into the $S=3/2$ Weyl
 point with the four-dimensional $\bar F$ representation
and the $S=1/2$ Weyl point 
with the two-dimensional $\bar E_2$ representation.

Here we note the magnitude of the SOC in the $K_4$ carbon system.
In Sec.\ \ref{sec:SOC}, we have analyzed 
   the SOC based on the tight-binding model.
It is well known that the SOC is small in the carbon system
because of the light atom.
We have performed the first-principles calculation 
to the $K_4$ carbon system, including the SOC.
We have verified that the energy splitting given in Fig.\ \ref{fig5}
 can be reproduced from the first-principles calculation, but 
the energy splitting of the sixfold degeneracy at the $H$ point
 is small $\sim$ 10 meV.
As has been discussed in Ref.\  \cite{Young:2012kz},
  the replacement of carbon atoms with heavier atoms 
  enlarges the energy splitting.
Further physical and chemical analyses are necessary for the realization of  
$S=3/2$ Weyl semimetal in the $K_4$ crystal, i.e., 
for clarifying the conditions 
that the $S=3/2$ Weyl
 point emerges at the Fermi energy without 
the other Fermi surfaces.

Finally we briefly discuss the relevance of the 
 present analysis to the recently synthesized $K_4$ crystal.
The first success in synthesizing the $K_4$ crystal was achieved
\cite{Mizuno:2015ip},
where the constituting component is 
a molecule (called the NDI-$\Delta$), instead of the carbon atom.
In this material,
the frontier molecular orbitals are extended along the
neighboring molecules, i.e., 
the inter-molecular overlapping is of the  $\sigma$ type.
In addition, the long distance hoppings are not relevant 
since the distance between the next-nearest-neighbor  molecules 
is large ($=\sqrt{3/8} a\sim 18$ \AA).
Thus the 
dispersive band structure is 
 similar to that shown in Fig.\ \ref{fig3}(a).
In addition, 
  the filling factor $n$
for the NDI-$\Delta$ system, 
was evaluated as $n\approx 1.4$ \cite{Mizuno:2015ip}, 
 which  is close to $n=4/3$.
This indicates a possibility that the Fermi energy lies on 
  the $S=1$ Dirac point, i.e., the $S=1$ Dirac semimetal. 
Additionally,   nontrivial flat bands have been pointed out 
 reflecting the peculiar molecular structure of NDI-$\Delta$
\cite{Mizuno:2015ip}.
Further theoretical investigation 
 needed for analyzing the electronic states in the 
newly-discovered $K_4$ crystal is left for future work.

\textit{Note added.}
We became aware of the paper by Ma\~nes \cite{Manes:2012fi}
at the final stage of this work, 
where the tight-binding Hamiltonian of a model
with the space group $I4_132$
had been considered explicitly.
 This is essentially identical to Eq.\
(\ref{eq:H_tb}), where the $S=1/2$ Dirac points at the $P$ point
and the  $S=1$ Dirac points at the $\Gamma$ and $H$ points 
had been pointed out.
The similar dispersion relation has been 
pointed out recently 
in the Kitaev spin model on 
the $K_4$ crystal \cite{Hermanns:2014tr,OBrien:2016bj},
where the dispersion is for the Majorana fermion.
We also became aware of the recent paper \cite{Bradlyn:2016cr}, which
 gave general arguments on the higher-spin Dirac or Weyl dispersions
and classified them by the space group symmetries.

\acknowledgments

The author is thankful for fruitful discussions with 
K. Awaga, M.\ M.\ Matsushita, Y.\ Shuku, A.\ Mizuno, R.\ Suizu, 
V.\ Robert, A.\ Kobayashi, and
A.\ Yamakage.
The author also thanks M. Hermanns and A. D. Zabolotskiy
for pointing out related references.
This work was supported by Grant-in-Aid for Scientific Research 
(24740232, 25400370, and 16K05442) 
from the 
Ministry of Education, Culture, Sports, Science and Technology, Japan,
and Japan-France Integrated Action Program, 
from Japan Society for the Promotion of Science.

\end{document}